\documentclass{ws-procs975x65}

\newcommand{\cD}{\ensuremath{\mathcal D} }
\newcommand{\cDbar}{\ensuremath{\overline{\mathcal D}} }
\newcommand{\Ibb}{\ensuremath{\mathbb I} }
\newcommand{\cN}{\ensuremath{\mathcal N} }
\newcommand{\cO}{\ensuremath{\mathcal O} }
\newcommand{\cP}{\ensuremath{\mathcal P} }
\newcommand{\cQ}{\ensuremath{\mathcal Q} }
\newcommand{\cU}{\ensuremath{\mathcal U} }
\newcommand{\cUbar}{\ensuremath{\overline{\mathcal U}} }
\newcommand{\al}{\ensuremath{\alpha} }
\newcommand{\eps}{\ensuremath{\epsilon} }
\newcommand{\la}{\ensuremath{\lambda} }
\newcommand{\lalat}{\ensuremath{\lambda_{\rm lat}} }
\newcommand{\ka}{\ensuremath{\kappa} }
\newcommand{\muhat}{\ensuremath{\widehat \mu} }
\newcommand{\UN}{\ensuremath{\mbox{U(}N\mbox{)}} }
\newcommand{\SUN}{\ensuremath{\mbox{SU(}N\mbox{)}} }
\newcommand{\Uone}{\ensuremath{\mbox{U(1)}} }
\renewcommand{\Re}{\ensuremath{\mbox{Re}\;} }
\renewcommand{\Im}{\ensuremath{\mbox{Im}\;} }
\newcommand{\lsim}{\ensuremath{\lesssim} }
\newcommand{\gsim}{\ensuremath{\gtrsim} }
\newcommand{\nn}{\nonumber}

\newcommand{\pf}{\ensuremath{\mbox{pf}\,} }
\newcommand{\Tr}[1]{\ensuremath{\mbox{Tr}\left[ #1 \right]} }
\newcommand{\vev}[1]{\ensuremath{\left\langle #1 \right\rangle} }
\newcommand{\eq}[1]{Eq.~\ref{#1}}
\newcommand{\fig}[1]{Fig.~\ref{#1}}
\newcommand{\secref}[1]{Section~\ref{#1}}
\def\figheight{7 cm}

\begin{document}
\title{Maximally supersymmetric Yang--Mills on the lattice}
\author{David Schaich and Simon Catterall}
\address{Department of Physics, Syracuse University, Syracuse, New York 13244, United States \\ E-mail: dschaich@syr.edu}

\begin{abstract} 
  We summarize recent progress in lattice studies of four-dimensional $\cN = 4$ supersymmetric Yang--Mills theory and present preliminary results from ongoing investigations.
  Our work is based on a construction that exactly preserves a single supersymmetry at non-zero lattice spacing, and we review a new procedure to regulate flat directions by modifying the moduli equations in a manner compatible with this supersymmetry.
  This procedure defines an improved lattice action that we have begun to use in numerical calculations.
  We discuss some highlights of these investigations, including the static potential and an update on the question of a possible sign problem in the lattice theory. \\
\end{abstract}

\bodymatter
\noindent In recent years there has been tremendous progress in lattice studies of four-dimensional $\cN = 4$ supersymmetric Yang--Mills (SYM) theory, employing a lattice formulation that exactly preserves a single supercharge $\cQ$.\cite{Kaplan:2005ta, Unsal:2006qp, Catterall:2007kn, Damgaard:2008pa, Catterall:2009it}
Advances both in the scale of computations and in the algorithms employed\cite{Schaich:2014pda} have provided the first ab~initio numerical results to be confronted with perturbative and holographic predictions for quantities such as the static potential.\cite{Catterall:2014vka, Catterall:2014vga}
At the same time this work has also led to improvements of the lattice construction, in particular the development of a new procedure to regulate flat directions in a manner compatible with the \cQ supersymmetry.\cite{Catterall:2015ira}
Given the central role of $\cN = 4$ SYM in the AdS/CFT correspondence that relates it to quantum gravity, it is important to continue large-scale lattice investigations of the theory away from the regime of weak coupling and for arbitrary numbers of colors $N$.
The recent progress, although significant, is only the beginning of this effort.

In this proceedings we briefly review the latest developments and present some new preliminary results from ongoing lattice $\cN = 4$ SYM studies.
We begin in the next section by summarizing the new procedure to regulate flat directions without breaking the exact \cQ supersymmetry.
Numerical calculations using the resulting improved lattice action exhibit dramatically reduced violations of supersymmetric Ward identities and much more rapid approach to the continuum limit.
We are now carrying out large-scale studies using this improved action.
In \secref{sec:potential} we revisit the static potential, checking that the improved action reproduces previous results.
The improved action has also allowed us to gain new insight into the possible sign problem of lattice $\cN = 4$ SYM, which we present in \secref{sec:pfaffian}.
We conclude with some discussion of other work currently underway and some of the next steps in our wide-ranging investigations.

\section{Supersymmetric deformation of the moduli space} 
A convenient starting point for lattice $\cN = 4$ SYM is the direct discretization of the continuum action produced by topological twisting,\cite{Marcus:1995mq, Kapustin:2006pk}
\begin{align}
  S_{\rm formal} & = S_{\rm exact} + S_{\rm closed}                                                                                                   \label{eq:S0} \\
  S_{\rm exact}  & = \frac{N}{2\lalat} \sum_n \Tr{\cQ \left(\chi_{ab}(n)\cD_a^{(+)}\cU_b(n) + \eta(n) \cDbar_a^{(-)}\cU_a(n) - \frac{1}{2}\eta(n) d(n) \right)} \nn \\
  S_{\rm closed} & = -\frac{N}{8\lalat} \sum_n \Tr{\eps_{abcde}\ \chi_{de}(n + \muhat_a + \muhat_b + \muhat_c) \cDbar_c^{(-)} \chi_{ab}(n)},                    \nn
\end{align}
with repeated indices summed.
All indices run from 1 through 5, requiring that the theory be formulated on the $A_4^*$ lattice with five linearly dependent basis vectors symmetrically spanning four space-time dimensions.\cite{Unsal:2006qp, Catterall:2014vga}
The five complexified gauge links $\cU_a$ contain both the gauge and scalar fields, and result in an enlarged $\UN = \SUN \otimes \Uone$ gauge invariance, where $N$ is the number of colors.

Both the SU($N$) and U(1) sectors necessarily possess flat directions that destabilize the vacuum of the lattice theory.
We recently introduced a general method that can be applied to regulate these flat directions in a manner compatible with the \cQ supersymmetry.\cite{Catterall:2015ira}
The procedure is to deform the $\cQ \left(\eta \cDbar_a^{(-)}\cU_a\right)$ term in the action, \eq{eq:S0}, as
\begin{equation}
  \label{eq:generic_deform}
  \cQ\; \Tr{\eta(n) \left(\cDbar_a^{(-)}\cU_a(n)\right)} \to \cQ\; \Tr{\eta(n) \left(\cDbar_a^{(-)}\cU_a(n) + G\cO(n)\Ibb_N\right)}
\end{equation}
where $G$ is a new tunable coupling and $\cO(n)$ is an appropriate gauge-invariant bosonic operator.
This $\cQ$-exact deformation modifies the auxiliary field equations of motion, $d(n) = \cDbar_a^{(-)}\cU_a(n) + G \cO(n)\Ibb_N$, which determine the moduli space of the system.
Since $\cQ\; \eta = d$, this also modifies the \cQ Ward identity
\begin{equation}
  \vev{\sum_n \Tr{\cQ\; \eta(n)}} = NG \vev{\sum_n \cO(n)} = 0,
\end{equation}
where the $\sum_n \Tr{\cDbar_a \cU_a(n)}$ term vanishes due to the sum over all lattice sites and the structure of the finite-difference operator $\cDbar_a$.\cite{Catterall:2007kn, Damgaard:2008pa}

Because the flat directions associated with the U(1) sector produce especially severe lattice artifacts,\cite{Catterall:2014vka, Catterall:2015ira} we take
\begin{equation}
  \cO(n) = \sum_{a \neq b} \left(\det\cP_{ab}(n) - 1\right) = 2\Re\sum_{a < b}\left(\det\cP_{ab}(n) - 1\right),
\end{equation}
where $\cP_{ab} = \cP_{ba}^*$ is the oriented plaquette in the $a$--$b$ plane.
The corresponding Ward identity is $\sum_n \sum_{a \neq b} \vev{\det\cP_{ab}(n) - 1} = 0$, implying $\vev{\Re\det\cP_{ab}} = 1$.
However, $\vev{\Im\det\cP_{ab}}$ is still unconstrained, and the SU($N$) flat directions still must be regulated as well.
We address both of these issues by adding to the action a soft \cQ breaking scalar potential,
\begin{equation}
  \label{eq:Ssoft}
  S_{\rm soft} = \frac{N}{2\lalat} \mu^2 \sum_n \sum_a \left(\frac{1}{N} \Tr{\cU_a(n) \cUbar_a(n)} - 1\right)^2.
\end{equation}

\begin{figure}[bp]
  \centering
  \includegraphics[width=0.425\linewidth]{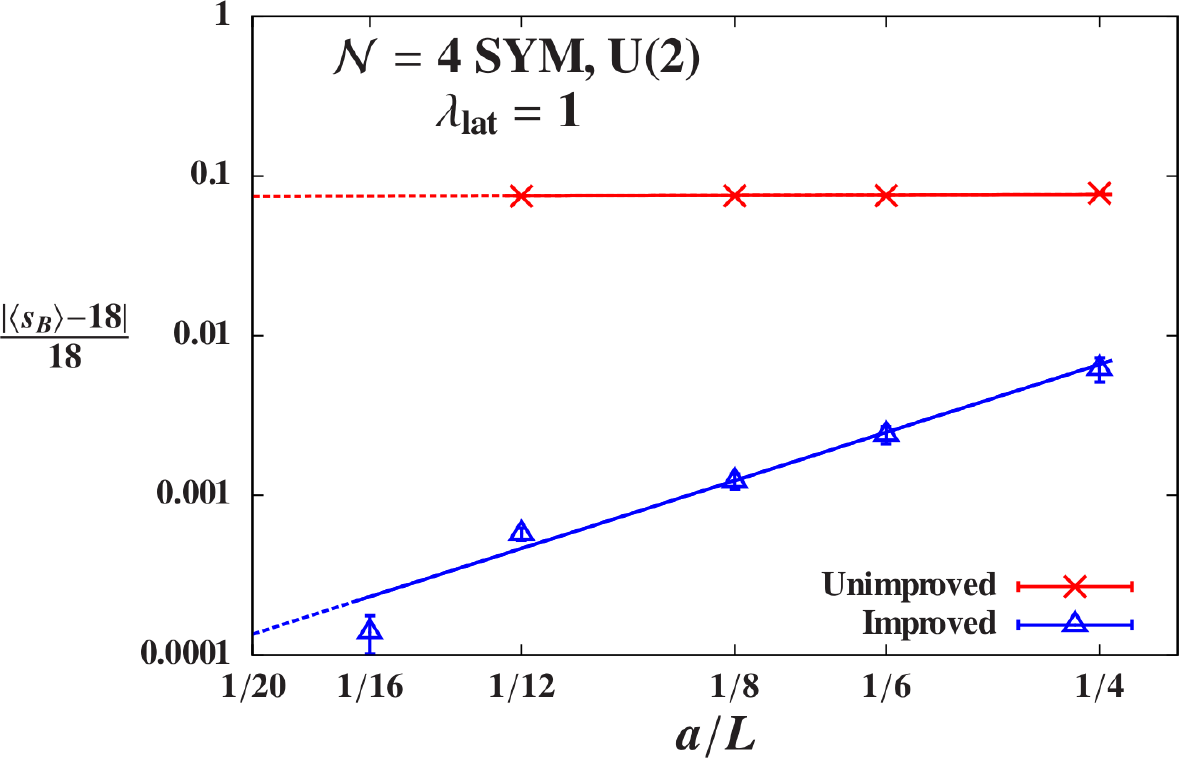}\hfill
  \includegraphics[width=0.475\linewidth]{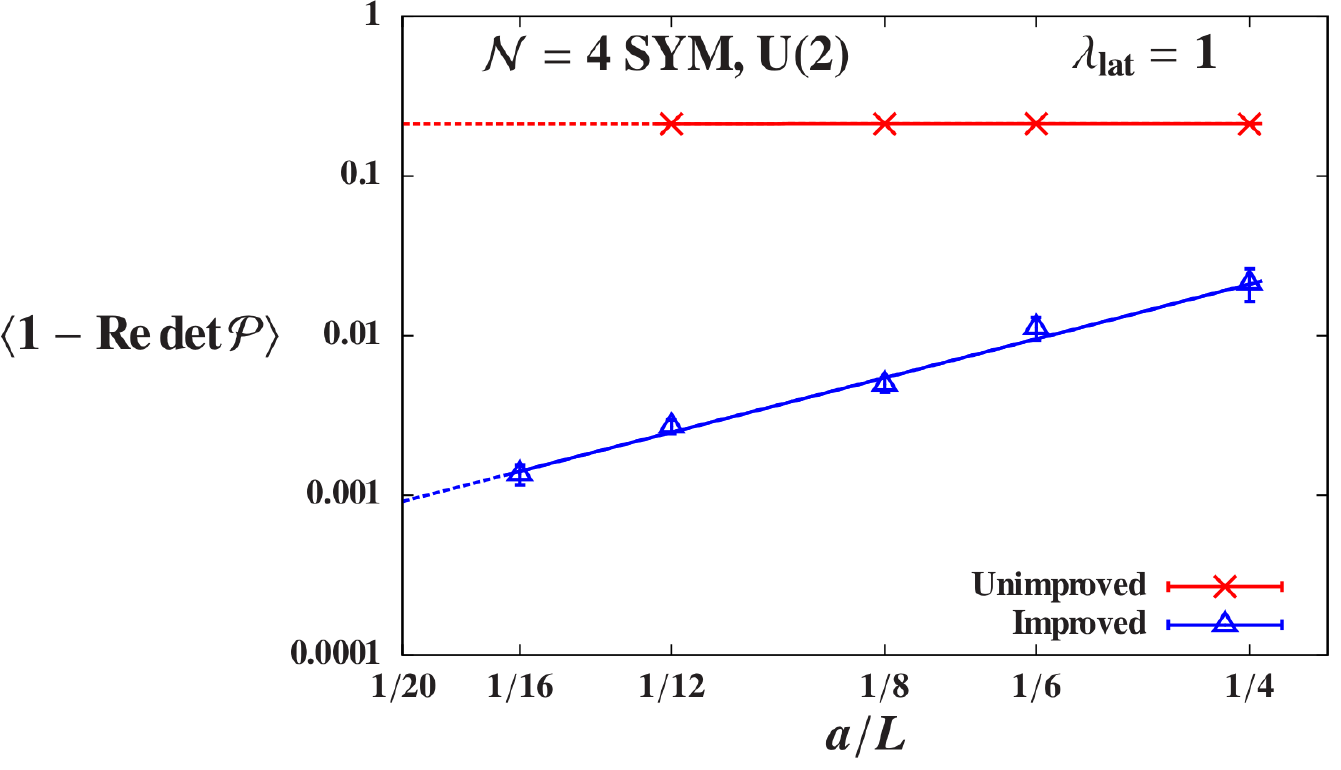}
  \caption{\label{fig:improvement}Continuum extrapolations of \cQ Ward identity violations on log--log axes with power-law fits, comparing the new improved lattice action\protect\cite{Catterall:2015ira} with the unimproved action used in previous works.\protect\cite{Catterall:2014vka, Schaich:2014pda, Catterall:2014vga}  Both deviations of the bosonic action from its exact supersymmetric value (left) and $\vev{1 - \Re\det\cP}$ (right) are much smaller for the improved action and approach the $a / L \to 0$ continuum limit much more rapidly, roughly $\propto (a / L)^2$.  For both actions we fix $\lalat = 1$ and the plaquette determinant coupling ($G$ or $\ka$), while scaling $\mu \propto 1 / L$.}
\end{figure}

Previous numerical studies used both this scalar potential as well as an additional soft \cQ breaking potential term that constrained the plaquette determinant.\cite{Catterall:2014vka, Schaich:2014pda, Catterall:2014vga}
\fig{fig:improvement} illustrates the improvement that results from our new method of lifting the U(1) flat directions.
Both plots in this figure quantify lattice artifacts by considering violations of \cQ Ward identities, which must vanish in the $a / L \to 0$ continuum limit.
The left plot addresses the Ward identity that fixes the value of the bosonic action per lattice site, $s_B = 9N^2 / 2$.
The right plot shows $1 - \vev{\Re \det\cP}$ as discussed above; while only the improved action constrains this through a Ward identity it must still vanish for the unimproved action as the U(1) sector decouples in the continuum limit.
In both cases the improved action produces much smaller lattice artifacts and approaches the continuum limit much more rapidly, roughly $\propto (a / L)^2$.
This scaling suggests that the improved action preserves the \cQ supersymmetry well enough to eliminate any significant effects of dimension-5 operators, all of which are forbidden by $\cQ$.\cite{Catterall:2015ira}

\section{\label{sec:potential}Revisiting the static potential} 
The main physical quantity considered by previous studies using the unimproved action was the static potential.
In our earlier work\cite{Catterall:2014vka, Catterall:2014vga} we found the static potential to be coulombic at both weak and strong coupling, $V(r) = A + C / r$ with vanishing string tension.
For gauge group U(2) our results for the Coulomb coefficient $C(\la)$ agreed with perturbation theory\cite{Pineda:2007kz, Stahlhofen:2012zx, Prausa:2013qva} for all $\la \lsim 2$ that we could investigate, where $\la \equiv \lalat / \sqrt{5}$ is the continuum-normalized bare 't~Hooft coupling.
For $N = 3$, however, we observed a significant departure from perturbation theory for $\la \gsim 1$, which could conceivably be the first sign that our results were approaching the famous large-$N$ prediction $C \propto \sqrt{\la}$ at strong coupling $1 \ll \la \ll N$.\cite{Rey:1998ik, Maldacena:1998im}

\begin{figure}[bp]
  \centering
  \includegraphics[width=0.45\linewidth]{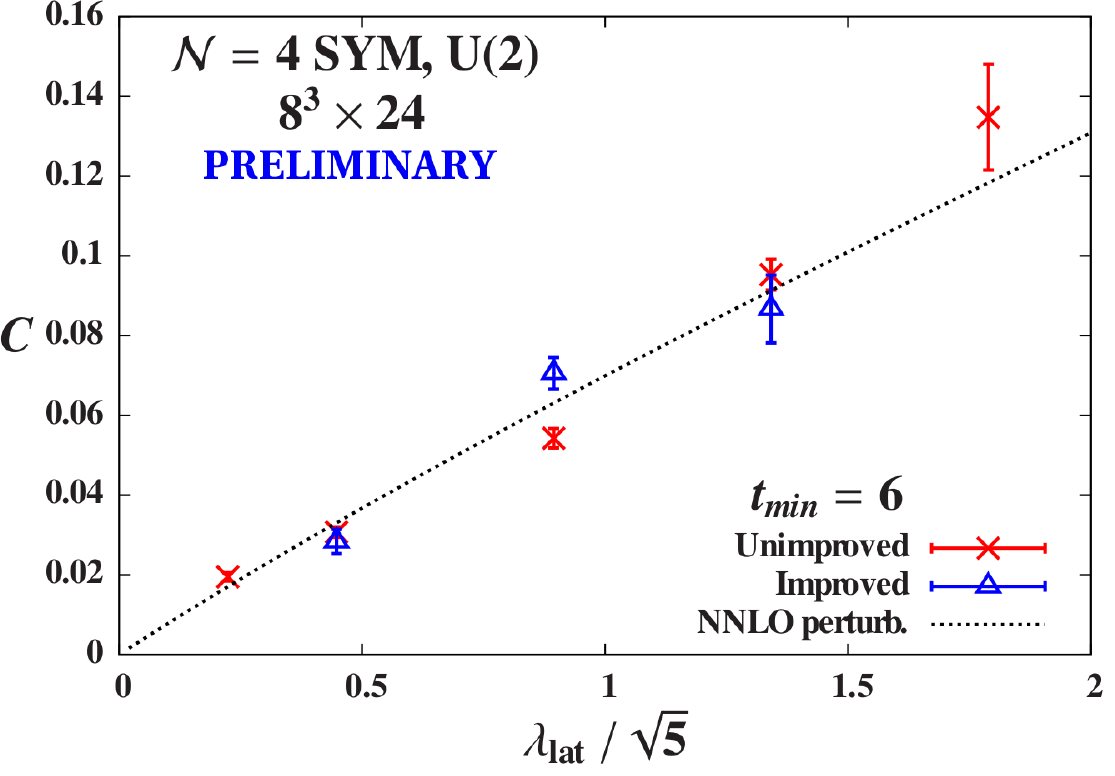}\hfill
  \includegraphics[width=0.45\linewidth]{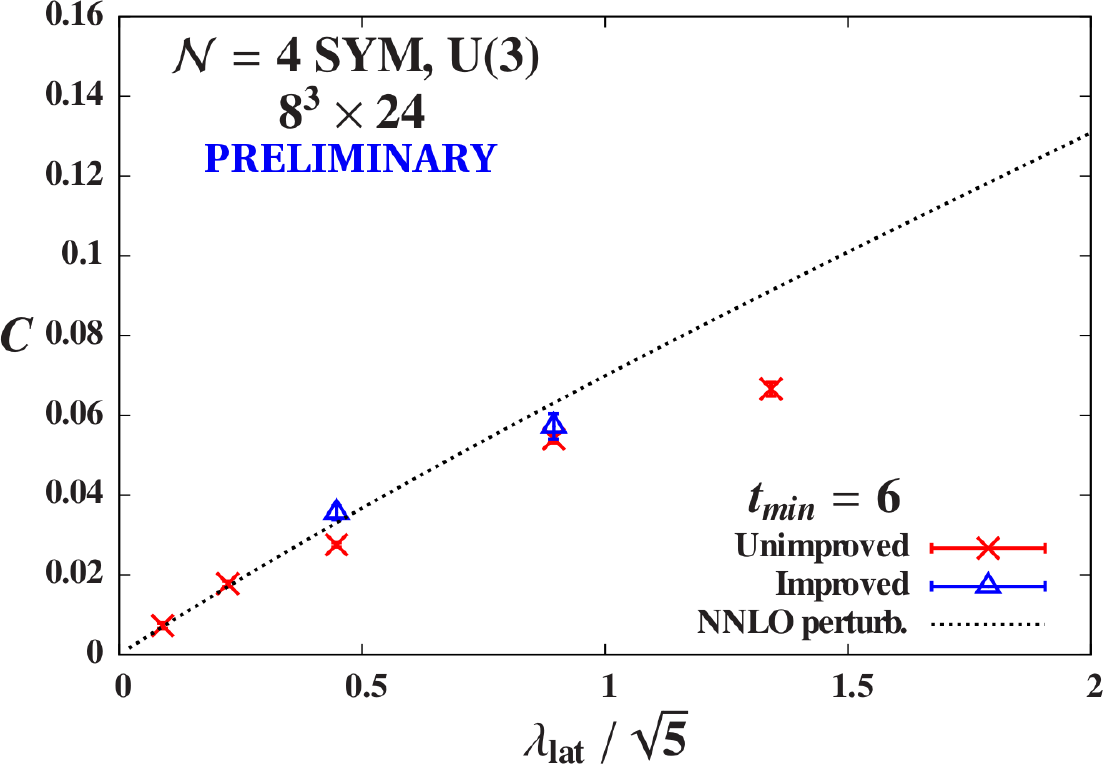}
  \caption{\label{fig:C}Static potential Coulomb coefficients $C$ vs.\ the bare 't~Hooft coupling $\lalat / \sqrt{5}$, comparing preliminary results obtained using the improved action with previously published unimproved results.\protect\cite{Catterall:2014vka, Catterall:2014vga}  For gauge group U(2) (left) we observe agreement with perturbation theory.  The $N = 3$ results (right) fall significantly below the NNLO perturbative prediction\protect\cite{Pineda:2007kz, Stahlhofen:2012zx, Prausa:2013qva} for $\la \gsim 1$.}
\end{figure}

We are currently repeating and extending these studies using the improved action that will reduce any effects of lattice artifacts.
As shown in \fig{fig:C}, our preliminary results from this ongoing work are consistent with the earlier observations summarized above.
Since the improved action allows us to consider significantly stronger couplings, we are in the process of exploring larger $\la$, as well as investigating $N = 4$.
Comparing results for all three gauge groups U(2), U(3) and U(4) will be extremely useful to identify systematic trends, but is numerically challenging\cite{Schaich:2014pda} due to the rapid growth of computational costs $\propto N^5$.

\section{\label{sec:pfaffian}New insight into the pfaffian phase} 
The improved action allows access to two new sources of information on the possible sign problem of lattice $\cN = 4$ SYM.
A sign problem is possible because gaussian integration over the fermion fields in the lattice path integral produces a pfaffian that is potentially complex, $\int \left[d\Psi\right] e^{-\Psi^T \cD \Psi} \propto \pf \cD = |\pf \cD| e^{i\al}$ where \cD is the fermion operator.
All our numerical calculations ``quench'' the phase $e^{i\al} \to 1$ in order to take advantage of efficient importance sampling algorithms.\cite{Schaich:2014pda}
For the unimproved action subsequent measurements of the pfaffian displayed no indication of a sign problem: the phase $e^{i\al}$ exhibited very small and volume-independent fluctuations around unity.\cite{Catterall:2014vka}

\begin{figure}[bp]
  \centering
  \includegraphics[height=\figheight]{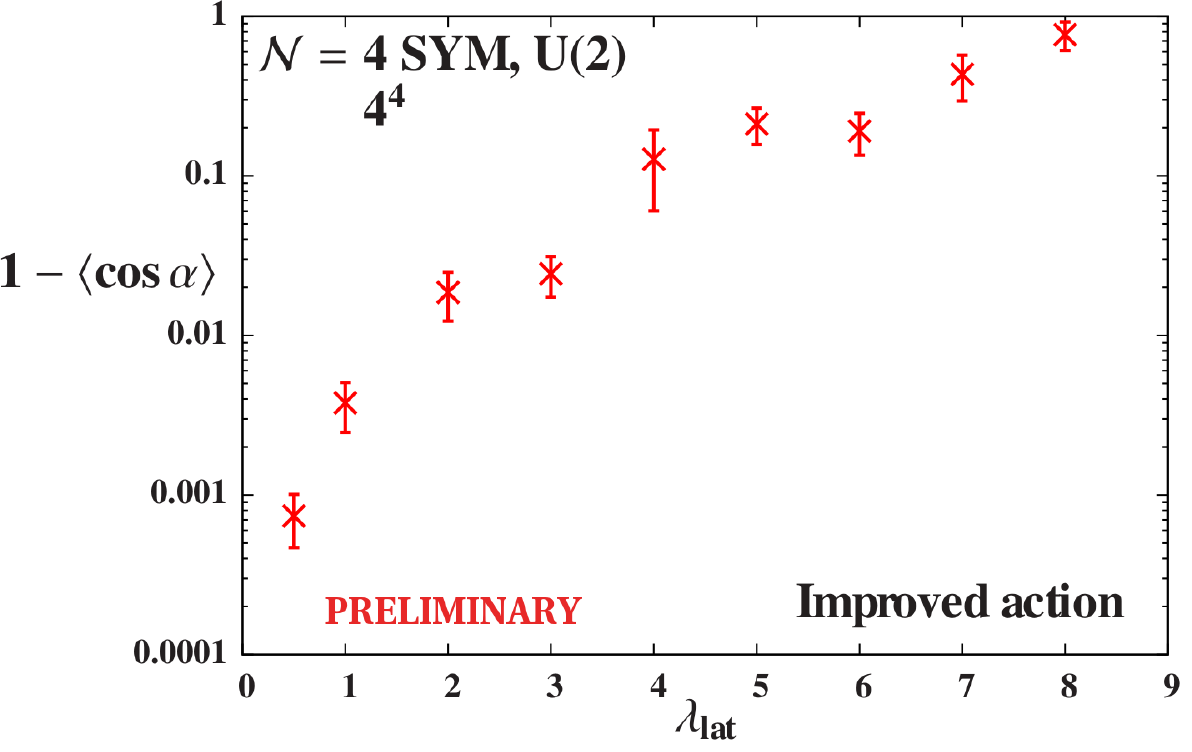}
  \caption{\label{fig:pfaffian}Dependence of the pfaffian phase fluctuations on the bare 't~Hooft coupling $\lalat$, for $4^4$ lattices and gauge group U(2).  $1 - \vev{\cos\al}$ vanishes when the pfaffian is real and positive, and its large value at strong coupling indicates significant fluctuations in the phase, i.e., an apparent sign problem.  For each \lalat we use the smallest acceptable $\mu$ and $G$ (Eqs.~\protect\ref{eq:Ssoft} and \protect\ref{eq:generic_deform}), both of which must increase with the coupling.  Larger \lalat and larger $G$ both appear to produce larger fluctuations in $e^{i\al}$, potentially compounding the sign problem at strong coupling.}
\end{figure}

However, most of our unimproved investigations of the pfaffian phase focused on relatively weak couplings, and for volumes larger than $3^3\times 4$ we considered only $\lalat = 1$.
As mentioned in the previous section, the improved action allows us to study significantly stronger couplings while still controlling lattice artifacts, and we are taking advantage of this to begin measuring $\vev{e^{i\al}}$ for larger $\lalat \leq 8$.
Preliminary results from $4^4$ lattices are shown in \fig{fig:pfaffian}, and reveal that fluctuations in the pfaffian phase increase significantly at stronger couplings $\lalat \gsim 4$, apparently indicating a sign problem.
Part of this increase may result from the larger values of the auxiliary coupling $G$ (\eq{eq:generic_deform}) that have to be used as \lalat increases;\cite{Catterall:2015ira} larger $G$ appears to be correlated with larger pfaffian phase fluctuations at fixed $\lalat$.

\begin{figure}[bp]
  \centering
  \includegraphics[height=\figheight]{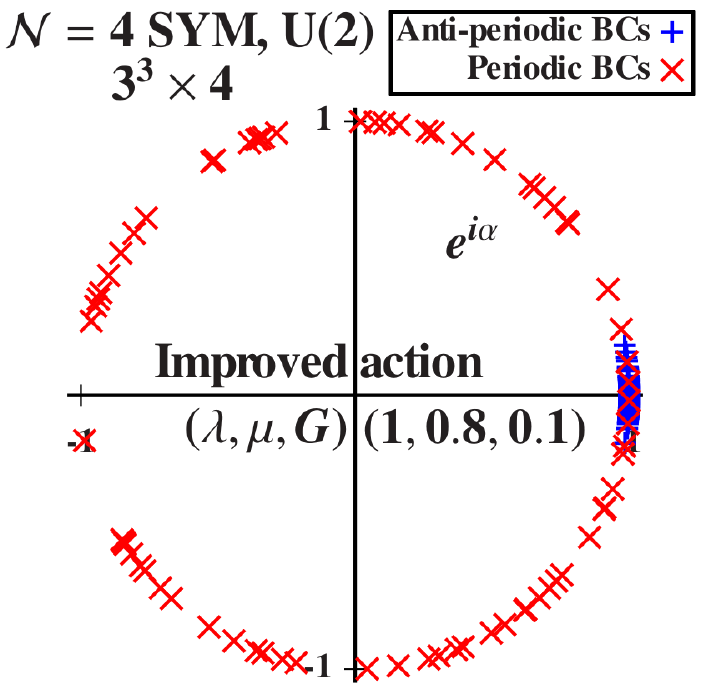}
  \caption{\label{fig:sign}Scatter plot of pfaffian phase measurements from independent $3^3\times 4$ ensembles using either periodic or anti-periodic temporal BCs for the fermion fields (and periodic BCs for all other fields and all other directions), with all couplings fixed.  While anti-periodic BCs produce $e^{i\al} \approx 1$, periodic BCs lead to uncontrolled fluctuations and $\vev{e^{i\al}}$ consistent with zero.}
\end{figure}

The other new development is related to the fact that the $\cQ$-invariant deformation in \eq{eq:generic_deform} breaks an $\eta \to \eta + c\Ibb_N$ shift symmetry and lifts the corresponding U(1) fermion zero mode.\cite{Catterall:2015ira}
The improved action therefore allows us to compute the pfaffian of \cD with fully periodic boundary conditions (BCs).
This is not possible for the unimproved lattice action, which relies on the fermions' anti-periodic (thermal) temporal BCs to lift this zero mode.
In \fig{fig:sign} we compare pfaffian phase measurements with either periodic or anti-periodic temporal BCs for the fermion fields and everything else fixed.
The contrast is dramatic: anti-periodic BCs produce $e^{i\al} \approx 1$ while fully periodic BCs lead to uncontrolled fluctuations and $\vev{e^{i\al}}$ consistent with zero.
It is not yet clear to us why the pfaffian is so sensitive to these BCs.
Even more mysteriously, all other observables change very little between these two ensembles, despite the apparent presence of a sign problem in one case.

\section{Work in progress and next steps} 
Significant work remains necessary to complete both of the ongoing studies discussed above.
We are still accumulating statistics for several points in each of Figs.~\ref{fig:C} and \ref{fig:pfaffian}.
For the static potential we are in the process of studying stronger couplings and gauge group U(4) to confirm the departure from perturbation theory and explore whether this behavior is consistent with holographic expectations.\cite{Rey:1998ik, Maldacena:1998im}
We are also improving our static potential data analysis, to account for short-distance discretization effects when fitting $V(r)$ and to compute renormalized couplings for more consistent comparisons with continuum predictions.

For the pfaffian phase, our new results at stronger 't~Hooft couplings need to be supplemented by measurements on larger volumes $V$, to see whether the volume-independence observed for the unimproved action persists with the improved action.
We also want to investigate $e^{i\al}$ for U($N$) gauge groups with larger $N = 3$ and 4.
Both of these goals are challenging;\cite{Schaich:2014pda} the computational costs of our pfaffian measurements increase $\propto N^6$ and $\propto V^3 \sim L^{12}$.
At the same time we continue to search for improved qualitative understanding of the system's insensitivity to apparent sign problems at strong coupling or with fully periodic BCs.

Of course there are many other interesting aspects of lattice $\cN = 4$ SYM that we lack the space to discuss in this proceedings.
One example is the restoration of the other fifteen supersymmetries $\cQ_a$ and $\cQ_{ab}$ in the continuum limit.
This can be investigated by measuring violations of discrete R~symmetries,\cite{Catterall:2013roa} as we previously explored with the unimproved action.\cite{Catterall:2014vka, Catterall:2014vga}
We are currently revisiting this study using the improved action, and will also address the related tuning of a single marginal operator that may be required to obtain the correct continuum limit.\cite{Catterall:2014mha, Catterall:2014vga}

Our highest priority, however, is to compute the anomalous dimension of the Konishi operator $\cO_K = \sum_I \Tr{\Phi^I \Phi^I}$, where $\Phi^I$ are the six scalar fields of $\cN = 4$ SYM.
Lattice predictions for this quantity will be complementary to those obtainable in perturbation theory,\cite{Beem:2013hha} from holography,~\cite{Gubser:1998bc, Gromov:2009zb} or via the conformal bootstrap program.\cite{Beem:2013qxa}
This work is advancing rapidly, exploiting both finite-size scaling and Monte Carlo renormalization group techniques,\cite{Catterall:2014mha} and we hope to report initial results in the near future.

\paragraph{Acknowledgments:} We thank Joel Giedt, Poul Damgaard, Tom DeGrand and Aarti Veernala for helpful conversations and continuing collaboration on lattice $\cN = 4$ SYM, as well as Julius Kuti, Valentina Forini and Rainer Sommer for useful discussions about the static potential.
We are grateful for the hospitality of the Aspen Center for Physics, supported by the U.S.~National Science Foundation under Grant No.~PHYS-1066293.
This work was supported by the U.S.~Department of Energy (DOE), Office of Science, Office of High Energy Physics, under Award Numbers {DE-SC}0008669 and {DE-SC}0009998. 
Numerical calculations were carried out on the HEP-TH cluster at the University of Colorado and on the DOE-funded USQCD facilities at Fermilab.

\raggedright
\bibliographystyle{ws-procs975x65}
\bibliography{SCGT15}
\end{document}